\documentclass[11pt,twoside]{article}
\usepackage{./asp2014}

\aspSuppressVolSlug
\resetcounters

\begin{document}

\title{The GALAH Survey and Galactic Archaeology in the next decade}
\author{Sarah~L.~Martell$^1$ 
\affil{$^1$School of Physics, University of New South Wales, Sydney, NSW 2052, Australia; \email{s.martell@unsw.edu.au}}}

\paperauthor{Sarah~L~Martell}{s.martell@unsw.edu.au}{0000-0002-3430-4163}{University of New South Wales}{School of Physics}{Sydney}{NSW}{2052}{Australia}

\begin{abstract}
The field of Galactic Archaeology aims to understand the origins and evolution of the stellar populations in the Milky Way, as a way to understand galaxy formation and evolution in general. The GALAH (Galactic Archaeology with HERMES) Survey is an ambitious Australian-led project to explore the Galactic history of star formation, chemical evolution, minor mergers and stellar migration. GALAH is using the HERMES spectrograph, a novel, highly multiplexed, four-channel high-resolution optical spectrograph, to collect high-quality ${\rm R} \sim 28,000$ spectra for one million stars in the Milky Way. From these data we will determine stellar parameters, radial velocities and abundances for up to $29$ elements per star, and carry out a thorough chemical tagging study of the nearby Galaxy. There are clear complementarities between GALAH and other ongoing and planned Galactic Archaeology surveys, and also with ancillary stellar data collected by of major cosmological surveys. Combined, these data sets will provide a revolutionary view of the structure and history of the Milky Way.
\end{abstract}

\section{Galactic Archaeology}
Galactic Archaeology is a highly detailed approach to near-field cosmology, using the structure and history of the Milky Way to build a systematic understanding of general processes in galaxy evolution. One of the main goals of Galactic Archaeology is to identify the remnants of ancient building blocks of the Galaxy that are now dispersed.  Chemical tagging, which follows the reasoning that stars that form together at the same time and place should have matching abundance patterns since they are forming from the same gas cloud, was first suggested by \citet{FBH02} as a way to search the present-day Galaxy for stars that originated together in these building blocks.

Large-scale Galactic Archaeology surveys like GALAH \citep{DS15}, Gaia-ESO \citep{GR12} and APOGEE (Apache Point Observatory Galaxy Evolution Experiment, Holtzman et al.~2015\nocite{HS15}) will enable powerful chemical tagging by acquiring large, homogeneous, high-quality sets of spectroscopic data, and using those to derive high-precision stellar parameters and abundances. At present, the precision of the parameters and abundances is still approaching its intended level, which limits the chemical tagging that can be carried out with the data. This is in part a result of the diversity of stars in each survey: stars with very different effective temperatures, surface gravities and compositions have very different atmospheres and emergent spectra, and require different and specialized approaches. Such a large data volume requires automated reduction and analysis, and in such a complex and degenerate problem, the intuition and experience of expert high-resolution spectroscopists can be difficult to translate into algorithmic form.

The feasibility and capabilities of chemical tagging have been studied from both observational and theoretical perspectives. The observational work (eg, Bensby, Feltzing \& Oey 2014\nocite{BFO14}; Adibekyan et al.~2012\nocite{AS12}; Nissen \& Schuster 2011\nocite{NS11}; De Silva et al.~2007\nocite{DS07}) has had a significant impact on our view of complexity and population behavior in the Galactic disk and halo. However, each data set has had important limitations: high-precision abundance determinations can only be done with high-quality data, and the assembly of large high-precision data sets takes a significant amount of time. This restricts both the number of stars and the volume of the Galaxy that is explored by these studies, and therefore limits the conclusions that can be drawn from carrying out chemical tagging in those data sets (eg, Mitschang et al.~2014\nocite{MDS14}). It is possible to construct a composite data set from the literature, as in \citet{MDS13} and \citet{TF12}, but this brings the disadvantage of inhomogeneity in data sources, analysis procedures, and abundance scale zeropoints, limiting the accuracy with which the abundance patterns can be compared.

Theoretical studies of chemical tagging have focused on two topics: the ability of chemical tagging to distinguish between stars from different formation sites, and what can be learned about the Milky Way from chemically tagged data sets. \citet{ML15} use numerical simulations of cluster dissolution and disk evolution to investigate whether dissolved star clusters in the disk can be identified purely on kinematic grounds, and find that the probability of success is strongly correlated to initial cluster mass. They suggest that for most initial star-forming groups, abundance information is necessary in addition to kinematics to trace stars back to their initial groups. 

The fundamental limit to our ability to differentiate chemically between formation sites is whether the abundance pattern of each site is truly unique. This is postulated in \citet{FBH02} but has not yet been demonstrated - somewhat circularly, it is only by doing a large-scale chemical tagging experiment that we can learn whether this central tenet is true. Recent work by \citet{TC15} implicitly assumes that it is, while pointing out that uniqueness may not be sufficient: stars originating in distinct star-formation events that have similar abundance patterns may be grouped together by chemical tagging if enough independent elemental abundances are not measured or if the measurement precision on the abundances is too low. \citet{RB14}, in a study on Solar twins, also highlight the importance of measuring nucleosynthetically independent elements for chemical tagging. Using a complex model of star formation over the history of the Milky Way, \citet{TC15} find that the slope and high-mass cutoff of the initial cluster mass function affects the "clumpiness" of the distribution in chemical space, implying that the former quantities can be reconstructed by measuring the latter. They then go on to make predictions about the number of formation sites that can be recovered in large GA surveys with different levels of chemical space sampling, sample size, and initial cluster mass function parameters, providing thoughtful input on survey strategy. 

\section{The GALAH Survey}
GALAH will determine stellar parameters and up to 29 elemental abundances for one million stars, primarily in the Galactic thin and thick disk. A justification for the million-star sample and an explanation of why the thick disk presents the richest opportunity for chemical tagging are presented in \citet{BH15}. GALAH data are taken with the HERMES spectrograph \citep{S14} at the 3.9m Anglo-Australian Telescope (AAT), a recently completed four-channel fiber-fed high-resolution (R$\sim 28,000$) optical spectrograph with bandpasses placed to optimize the amount of information collected on stellar parameters and abundances.\footnote{Further details on HERMES can be found at http://www.aao.gov.au/science/instruments/current/HERMES} HERMES is located in a temperature-stabilised room at the AAT and is fed by a fiber cable attached to the 2dF fiber positioner \citep{LC02}, which sits at prime focus of the AAT and can place up to 392 science fibers in a two-degree-diameter field. HERMES produces data with a signal-to-noise ratio (SNR) of 100 per resolution element for stars with $V=14$ in one hour of exposure time under typical seeing conditions.

The main driver for GALAH survey strategy is a simple selection function, so that trends in the underlying Galactic populations can be inferred in a straightforward way from the behavior of the observed sample. As a result, GALAH considers all stars with $V \leq 14.3$, $\delta \leq +10$ and $|b| \geq 10$ that meet key 2MASS quality requirements to be valid survey targets. This initial catalogue contains 8,149,211 stars, which are tiled into 4303 fixed fields. Each GALAH survey observation includes 400 targets from one of these fields (there are several unique sets of targets in dense fields) with $12<V<14$, with 2dF configured as efficiently as possible using AAO's Configure software\footnote{Configure, and documentation on how it works, are available online at http://www.aao.gov.au/science/software/configure}. This typically results in 360 of those stars being allocated to a fiber, and standard survey exposure times are one hour, extended to 80 or 120 minutes in poor seeing.

The initial input catalogue has been extended to include stars in the Kepler-2 \citep{HS14} ecliptic campaign regions, since proposals for Galactic Archaeology-focused target selection in Kepler-2 (led by members of the GALAH survey team) have been very successful. In GALAH Kepler-2 fields the main goal is deriving stellar parameters and metallicities, and not detailed abundances, so targets as faint as 15.3 have acceptable SNR in a one-hour integration. GALAH has been observing since February 2014, and through April 2015 has observed 117,173 stars in 262 fields. Figure \ref{galaxia} shows predictions for the distribution of age, distance and [Fe/H] in the full GALAH survey sample, as predicted with the Galaxia program \citep{S11}.

\section{Survey synergies}
The three major ongoing Galactic Archaeology surveys, GALAH, APOGEE and Gaia-ESO, are observing complementary regions of the Galaxy. This reflects the locations of the observatories where they collect their data and the capabilities of their instruments, as well as differing science priorities. Because of limits on apparent magnitude and field density, GALAH observes primarily disk dwarfs, and the final data set will contain a thorough sample of stars within $\sim5$kpc of the Sun. APOGEE observes in the near-infrared, and as a result is able to observe stars at very low Galactic latitude \citep{ZJ13}, with additional pencil-beam observations of red giants in the halo and targeted observations of star clusters and the bulge. Gaia-ESO is a combination of a halo survey and a star cluster survey, so its sample covers a wide range in stellar age and metallicity, and it covers a significantly larger volume in the Milky Way than GALAH.

\subsection{Gaia}
The Gaia satellite \citep{P12} provides a tremendously important synergy for all studies of Galactic structure, dynamics and stellar populations. It will determine parallaxes and proper motions for $10^{9}$ stars in the Milky Way, to a magnitude limit of $V = 20$. Gaia will determine distances and 3D space velocities for all of these stars, providing an absolute baseline for stellar physics that has never before been available. Stars that GALAH will observe are the bright end of the Gaia sample, meaning that they will have the most precise results, with errors on parallax and proper motion of $10$ microarcseconds per year, translating to $1\%$ distance errors at $1$kpc and $0.7$kms$^{-1}$ velocity errors at 15 kpc.

Gaia distances can be used in conjunction with chemical tagging to test whether stars with similar abundance patterns formed at the same time. By constructing color-magnitude diagrams with absolute magnitudes, it will be straightforward to test whether stars that form a group in chemical abundance space also follow a single isochrone with an appropriate [Fe/H] and [$\alpha$/Fe]. This will serve as both a step to reject group non-members and a strong endorsement of the validity of groups identified in chemical space. Figure 2 in \citet{MDS14} shows an example of chemically tagged groups confirmed by both isochrone fitting and UVW velocities.

\subsection{Ongoing Galactic Archaeology surveys}
The SDSS-III APOGEE survey took data at Apache Point Observatory in the United States from mid-2011 until mid-2014, obtaining spectra for over 150,000 stars. Optical fibers carry light from the SDSS plug plates to the cryogenically cooled APOGEE spectrograph, allowing simultaneous acquisition of R$\sim22,500$ spectra for 300 stars across a seven-square-degree field. APOGEE's unique strength is observing in the near-infrared ($1.51\mu - 1.70\mu$), which allows observations of stars at any Galactic latitude. This permits studies that have not previously been possible, including large {\it in situ} studies of bulge stars and measurements of vertical and radial gradients in the disk that include stars all the way to the disk midplane (eg, Hayden et al. 2015\nocite{HB15}; 2014\nocite{HH14}). 

The SDSS-IV project includes two important extensions of APOGEE, referred to collectively as APOGEE-2: a continuation of the Northern survey through mid-2020, and a parallel Southern survey from mid-2016 through mid-2020 using a clone of the original spectrograph to be constructed and installed at the Ir\'en\'ee du Pont telescope at Las Campanas Observatory in Chile. The Southern extension will provide a dramatic improvement to APOGEE's ability to observe the bulge, and will also allow for focused observations in the Large and Small Magellanic Clouds.

APOGEE target selection emphasizes red giants, in order to maximize the spatial reach of the data set. As in the SDSS-II SEGUE and SDSS-III SEGUE-II surveys, the observing pattern is highly structured and relatively sparse on the sky.\footnote{Maps of the APOGEE observing footprint can be found at http://www.sdss.org/dr12/} As a result, APOGEE and GALAH data can be combined very effectively: in the disk, APOGEE has more distant stars, while GALAH will have an extremely thorough local sample, and a combined data set would have both high spatial resolution locally and a large spatial extent. In the halo, GALAH samples high-proper-motion nearby halo stars, while APOGEE observes an in situ population. A combined study would be able to test for radial trends and also contrast inner- and outer-halo populations.

The Gaia-ESO survey has been observing at the ESO Very Large Telescope in Chile since the end of 2011, using the FLAMES \citep{P00} facility to feed fibers to the GIRAFFE and UVES \citep{D00} spectrographs. To date Gaia-ESO has observed roughly half of its goal of 100,000 stars. Up to 130 targets in each field are observed multiple times with different GIRAFFE settings to obtain wavelength coverage across all spectral regions of interest at a resolution of R$\sim26,000 - 32,000$, and up to 8 bright stars in the field of view are simultaneously observed at higher resolution (R$\sim 47,000$) with UVES. 

Gaia-ESO target selection emphasizes membership in different Galactic components: K giants in the bulge, F stars in the halo and thick disk, stars in known halo streams, F and G stars in the Solar neighbourhood, and stars across the evolutionary sequence selected from previous studies of open clusters. The best combinations of GALAH and Gaia-ESO data will combine local and distant stars in the same component, similar to the GALAH-APOGEE combinations, or compare the unbiased GALAH sample in the Solar neighbourhood with the UVES sample in the same volume, which is specifically selected for F and G stars.

GALAH, APOGEE and Gaia-ESO are planning to co-observe in a number of fields. Placing the different surveys' results on the same abundance scale is essential for using them together, and observations across parameter space (stellar mass, composition and evolutionary phase) are critical for this survey cross-calibration. We can do this in a traditional way by looking for zeropoint offsets and trends between the results of the data analysis pipelines for stars observed by multiple surveys. In addition, we can use machine learning methods to cross-calibrate the surveys and potentially to identify information within the spectra themselves that the classical-analysis-based survey analysis pipelines overlook. The method developed by \citet{NH15} takes a training set of spectra with well-known parameters and abundances and assigns parameter and abundance values to other spectra based on their morphological similarities to the known spectra. Given appropriate crossover training sets, this will be a very powerful way to combine data from many different surveys, even those with very different data characteristics.

\section{Galactic Archaeology in the next decade}
The current Galactic Archaeology surveys will complete their observing campaigns in a few years. While a complete chemical tagging of the Galactic disk will require very large data sets, there are many scientific questions that the projects are already addressing. These include mapping the present-day structure of the disk \citep{HB15}; studies of stars in young clusters \citep{SJ15}, which address the core of the concept of chemical tagging; chemically unusual stars \citep{MR14}, which could be signals of minor mergers; comparisons of spectroscopic and asteroseismic stellar parameters \citep{EE14}; halo stars with characteristic globular cluster abundance patterns \citep{LK15}, which indicate {\it in situ} star formation in the halo; mapping Galactic structure and kinematics with diffuse interstellar bands (Puspitarini et al. 2015\nocite{PL15}; Zasowski et al. 2015\nocite{ZM15}); very metal-poor stars in the inner Galaxy (Howes et al. 2014\nocite{HA14}; Garcia Perez et al. 2013\nocite{GP13}), and studies of the abundance behaviour in globular and open star clusters (eg, Donati et al. 2015\nocite{DC15}; Tautvai{\v s}ien{\.e} et al. 2015\nocite{TD15}; Meszaros et al. 2015\nocite{MM15}). 

There are larger Galactic Archaeology surveys that are currently in the planning phase, with purpose-built instruments and dedicated observing facilities. These will be carried out as components of larger multipurpose projects: for example, the WEAVE \citep{DT12} survey science goals incorporate Galactic Archaeology, galaxy evolution and baryon acoustic oscillations, and the 4MOST \citep{DJ12} project will include sub-surveys on AGN and galaxy clusters in addition to Galactic Archaeology. 

The Galactic surveys planned by WEAVE and 4MOST have excellent complementarity, with similar data characteristics (R$\sim 5000$ and R$\sim 20,000$, broad optical wavelength coverage, samples on the order of millions of stars) and telescopes located in opposite hemispheres. WEAVE is on track to begin observations at the William Herschel Telescope on La Palma in 2017, and will use dual fiber positioning robots to allocate 1000 fibers across a two-degree-diameter field of view. These fibers can feed either a low- or high-resolution spectrograph, and WEAVE will use both for its stellar surveys. 4MOST will observe from the VISTA telescope at Cerro Paranal in Chile starting in 2018, and will use an Echidna-style fiber positioner \citep{GM00}. 4MOST will use its two spectrographs simultaneously, with 800 fibers dedicated to the high-resolution spectrograph and the remaining 800-1600 fibers feeding the low-resolution spectrograph. The new instrumentation developed for these two surveys will allow them to collect high-quality spectra for several million stars across the Milky Way.

Together these two projects will provide a dramatic step forward from the current generation of Galactic Archaeology surveys, in terms of the volume of the Galaxy studied and the thoroughness of the sampling. Their stellar surveys are designed to include stars down to the faint end of the Gaia catalogue, which will be published in full during their observing campaigns. These surveys will provide a revolutionary view of the detailed structure in the disk and halo.

Even survey projects that do not have stellar or Galactic astronomy as their main purpose will be important for Galactic Archaeology in the future. As an example, the primary science goals of the DESI (Dark Energy Spectroscopic Instrument, Levi et al. 2013\nocite{L13}) project are focused on baryon acoustic oscillations and redshift-space distortions. However, DESI expects to observe some ten million Milky Way stars during ancillary and bright-time observations in its five years of observing. Although the spectral resolution will be lower than the Galactic Archaeology-specific surveys (R$\sim 1500 - 4000$), this will be a tremendously powerful dataset for exploring the structure of the Milky Way, with comprehensive sampling and a sample that reaches to the furthest halo stars, providing a rich data set for data mining and detailed followup study.

\articlefigure{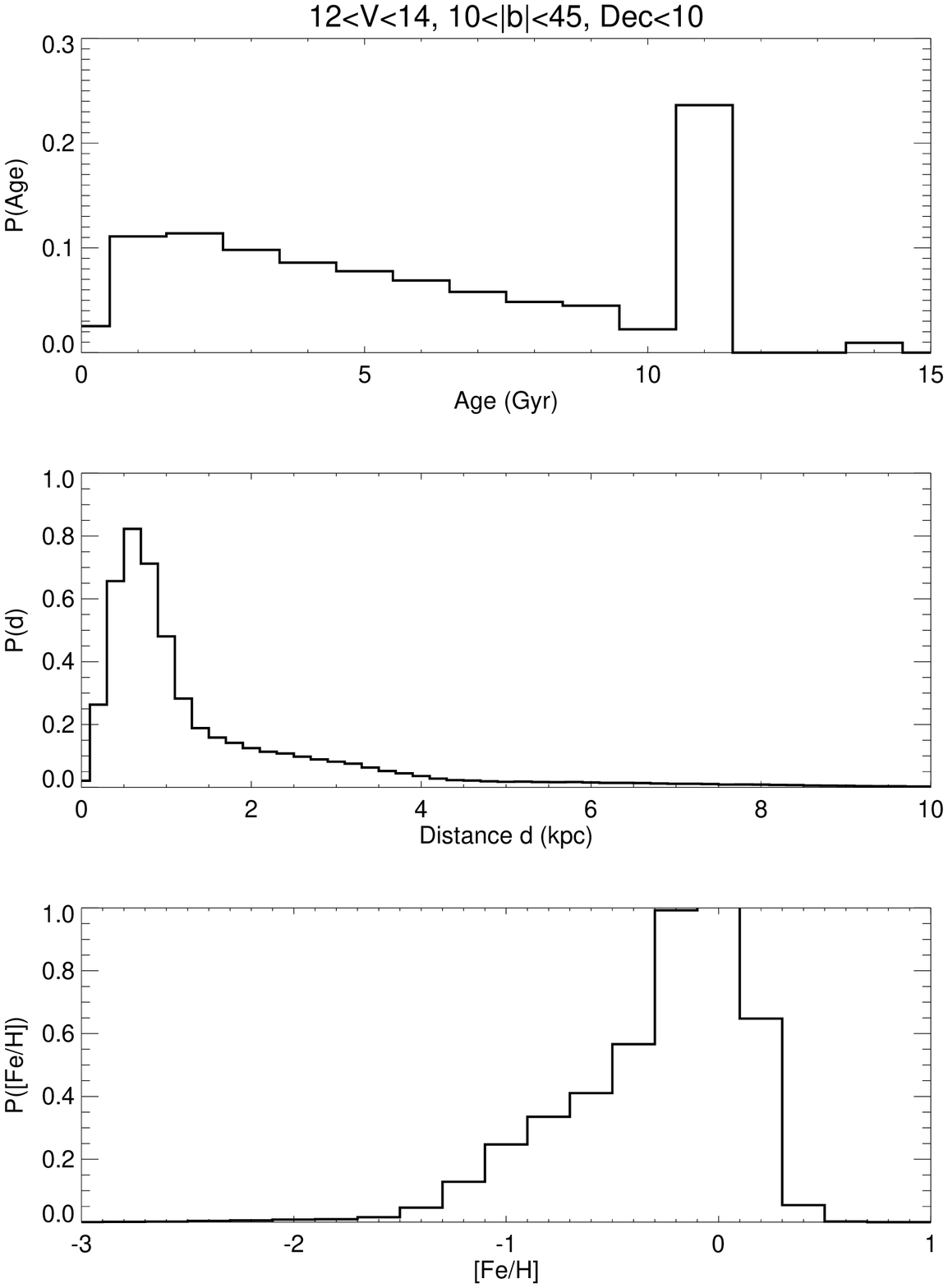}{galaxia}{Galaxia predictions for the properties of the GALAH observational sample}

\acknowledgements SLM acknowledges funding support for this project from Australian Research Council DECRA Fellowship DE140100598, and also acknowledges the efforts and expertise of the GALAH Survey team.

\bibliography{SMartell}  

\end{document}